\documentclass[reprint,aps,]{revtex4-1}

\usepackage{graphicx}
\usepackage{dcolumn}
\usepackage{bm}
\usepackage{amssymb}
\usepackage{amsfonts}
\usepackage{latexsym}
\usepackage{pdfsync}
\usepackage[T1]{fontenc} 
\usepackage{epsf,amsfonts}
\usepackage{epsfig}
\usepackage{amsthm}
\usepackage{amsmath}
\usepackage{color}
\usepackage{slashed}
\usepackage{mathrsfs}
\usepackage{float}
\usepackage{color}
\usepackage{esvect}
\usepackage{mathtools}
\usepackage{bbm}

\newcommand{\VEV}[1]{\left\langle #1\right\rangle}

\newcommand{\p}{\partial}

\newcommand{\MeV}{\;\text{MeV}}

\begin{document}

\title{A holographic study on QCD phase transition and phase diagram with two flavors} 

\author{Xin-Yi Liu}
\email{liuxinyi23@mails.ucas.ac.cn}
\affiliation{Department of Applied Physics, School of Physics and Electronics, Hunan University, Changsha 410082, China}
\affiliation{School of Fundamental Physics and Mathematical Sciences, Hangzhou Institute for Advanced Study, UCAS, Hangzhou 310024, China}
\author{Xiao-Chang Peng}
\email{pengxc1616@mails.jlu.edu.cn}
\affiliation{Department of Applied Physics, School of Physics and Electronics, Hunan University, Changsha 410082, China}

\author{Yue-Liang Wu}
\email{ylwu@itp.ac.cn}
\affiliation{School of Physical Sciences, University of Chinese Academy of Sciences, Beijing 100049, China}
\affiliation{CAS Key Laboratory of Theoretical Physics, Institute of Theoretical Physics, Chinese Academy of Sciences, Beijing 100190, China}
\affiliation{International Centre for Theoretical Physics Asia-Pacific (ICTP-AP), University of Chinese Academy of Sciences, Beijing 100049, China}

\author{Zhen Fang}
\thanks{zhenfang@hnu.edu.cn}
\affiliation{Department of Applied Physics, School of Physics and Electronics, Hunan University, Changsha 410082, China}
\affiliation{Hunan Provincial Key Laboratory of High-Energy Scale Physics and Applications, Hunan University, Changsha 410082, China}


\begin{abstract}
   We investigate the chemical potential effects of the equation of state and the chiral transition in an Einstein-Maxwell-dilaton-scalar system, which is obtained from an improved soft-wall AdS/QCD model coupled with an Einstein-Maxwell-dilaton system. The equations of state obtained from the model are in quantitative agreement with the lattice results at both zero and nonzero chemical potentials. The sensible chiral transition behaviors can be realized in the model. The QCD phase diagram with a CEP has also been obtained from the model.
\end{abstract}

\maketitle

\section{Introduction}\label{introduce1}

The investigations of Quantum Chromodynamics (QCD) phase transition and the construction of the QCD phase diagram are of paramount importance in the field of theoretical physics. These inquiries delve into the fundamental nature of matter and the universe, aiming to unravel the intricate behavior of quarks and gluons that make up protons, neutrons, and other hadrons. Understanding the QCD phase transition, which occurs under extreme conditions of temperature and baryon chemical potential, such as those in the early universe or within neutron stars, is critical for our comprehension of the fundamental forces governing the cosmos \cite{Aoki:2006we}. The QCD phase diagram, on the other hand, provides a comprehensive map of the different phases of nuclear matter and offers essential insights into the behavior of matter at various temperatures and densities. These investigations not only deepen our understanding of the building blocks of matter but also have practical applications, ranging from high-energy physics to astrophysics, impacting the way we perceive the universe and its evolution.

One of the key issues is to study the nature of QCD phase transition that takes place in hot and dense environments. QCD is known to exhibit two distinct types of phase transitions in these exteme conditions, namely the chiral and the deconfinement phase transitions. At low temperature $T$ and small baryon chemical potential $\mu_{B}$, QCD matter predominantly exists as confined hadrons due to strong quark confinement, with a nonzero chiral condensate that contributes to the hadron mass. As temperature rises, QCD matter will finally enter a deconfined phase of quark-gluon plasma (QGP) through a smooth crossover, with the chiral condensate approaching zero and the chiral symmetry restored \cite{Aoki:2006we,Bazavov:2011nk,Bhattacharya:2014ara}. It is commonly believed that the phase transition between hadronic matter and quark-gluon plasma changes from a smooth crossover to a first-order phase transition with the increase of $\mu_{B}$. Hence, the existence and properties of the critical end point (CEP) of the first-order transition line are essential for understanding the behavior of QCD matter under extreme conditions. The location of CEP depends on factors like the number of quark flavors and the strength of interactions. In extremely dense conditions, such as in the core of neutron stars, quark matter may undergo color superconductivity and exhibits novel phases \cite{Alford:2007xm}.

Efforts to understand the QCD phase transition and to construct the QCD phase diagram have been a focal point of both theoretical and experimental researches in the field of nuclear and particle physics. In the Large Hadron Collider (LHC) at CERN and the Relativistic Heavy Ion Collider (RHIC) at Brookhaven National Laboratory, high-energy nuclear collisions have been conducted to recreate conditions similar to the early universe, allowing us to investigate the properties of QCD matter under extreme conditions. Neutron star astrophysics and gravitational wave observations also  offer indirect insights into the behavior of dense nuclear matter under extreme conditions. Theoretical attempts at studying the QCD phase transition involve lattice QCD simulations \cite{Laermann:2003cv,Fukushima:2013rx} and  model calculations, such as the chiral effective model and various other theoretical frameworks \cite{Fischer:2009wc,Braun:2009gm,Qin:2010nq,Braun:2009gm,Son:2000xc,Ratti:2005jh,Schaefer:2007pw}. However, almost all the methods come with shortages and difficulties related to the non-perturbative nature of low-energy QCD. For the lattice QCD, the infamous sign problem exists when addressing issues at finite baryon density, even though various systematic schemes have been constructed to extrapolate the lattice results to finite baryon chemical potentials \cite{Allton:2002zi,JETSCAPE:2020shq,Borsanyi:2021sxv}.

In the last two decades, the gauge/gravity duality, also known as the AdS/CFT correspondence \cite{Maldacena:1997re,Gubser:1998bc,Witten:1998qj}, has been established as a promising approach to solve various strong-coupling related problems, especially the nonperturbative aspects of QCD \cite{Kruczenski:2003uq, Sakai:2004cn,Sakai:2005yt}. Indeed, it has been a long desire to construct a holographic dual of QCD, which is called the AdS/QCD program, aiming to provide quantitative descriptions for the low-energy QCD properties that is intrinsically non-perturbative. This includes the study of the hadron spectrum, the thermodynamic properties and also the phase structure of QCD. Many holographic QCD models have been constructed by a bottom-up approach, in consideration of the basic features of low-energy QCD, such as the spontaneous chiral symmetry breaking and the linear confinement property \cite{deTeramond:2005su,DaRold:2005mxj,Erlich:2005qh,Karch:2006pv,Csaki:2006ji,Cherman:2008eh,Fujita:2009wc,Fujita:2009ca,Colangelo:2009ra,Colangelo:2011sr,Li:2012ay,Li:2013oda,Shuryak:2004cy,Brodsky:2014yha,Tannenbaum:2006ch,Policastro:2001yc,Cai:2009zv,Cai:2008ph,Sin:2004yx,Shuryak:2005ia,Nastase:2005rp,Nakamura:2006ih,Sin:2006pv,Janik:2005zt,Herzog:2006gh,Gursoy:2007cb,Gursoy:2007er,Gherghetta:2009ac,Kelley:2010mu,Sui:2009xe,Sui:2010ay,Cui:2013xva,Cui:2014oba,Fang:2016uer,Fang:2016dqm,Herzog:2006ra,Li:2014hja,Li:2014dsa,Fang:2016cnt,Evans:2016jzo,Mamo:2016xco,Dudal:2016joz,Dudal:2018rki,Ballon-Bayona:2017dvv,Chen:2018msc}. The well-known ones include the hard-wall and soft-wall models \cite{DaRold:2005mxj,Erlich:2005qh,Karch:2006pv}, the modified versions of which could produce the light hadron spectra in agreement with measurements, and could also realize the proper chiral transition behaviors in the case of both two and $2+1$ flavors \cite{Chelabi:2015gpc,Chelabi:2015cwn,Fang:2016nfj,Li:2016smq,Bartz:2017jku,Bartz:2016ufc,Fang:2018vkp,Fang:2018axm,Fang:2019lmd}. The equation of state and the associated deconfinement phase transition at $\mu_{B}=0$ have been studied in the holographic framework by an Einstein-dilaton system with a suitable dilaton potential \cite{Gubser:2008yx,Gubser:2008ny,Gursoy:2008bu,Noronha:2009ud,Finazzo:2013efa,Finazzo:2014zga,Andreev:2009zk,Yaresko:2013tia,Yaresko:2015ysa,Colangelo:2010pe,Li:2011hp,He:2013qq,Yang:2014bqa,Fang:2015ytf,Rougemont:2017tlu,Li:2017ple,Zollner:2018uep,ChenXun:2019zjc}, while the Einstein-Maxwell-dilaton (EMD) system has been used to investigate the properties of QCD phase transition at $\mu_{B}\neq 0$ and to construct the QCD phase diagram \cite{DeWolfe:2010he,DeWolfe:2011ts,Critelli:2017oub}.
However, the chiral and deconfinement phase transitions, as two sides of one coin, should be realized simultaneously in a consistent way. For that, some efforts have been made to characterize these two kinds of phase transitions in a single holographic framework \cite{Jarvinen:2011qe,Alho:2012mh,Alho:2013hsa,Fang:2019lsz,Li:2022erd}.

In this work, we attempt to construct a holographic QCD model with the aim to provide a proper description for both the equation of state and the chiral transition at finite $\mu_{B}$ in the two-flavor case. The matter part of this model will be given by an improved soft-wall model which could generate spontaneous chiral symmetry breaking and realize the right chiral transition properties, at least qualitatively. The bulk background will be given by an EMD system which could describe the equation of state and the expected deconfinement properties of QCD. By integrating these two sectors into an Einstein-Maxwell-dilaton-scalar (EMDS) system, we find that the properties of deconfinement and chiral transitions could be characterized consistently in this holographic framework. Furthermore, we find that the model results of the equation of state at finite $\mu_{B}$ are in good agreement with the lattice results of two flavors, and the QCD phase diagram containing a CEP can also be obtained. After fixing the model parameters, we also investigate the equation of state in the reduced EMD system with the coupling $\beta=0$, and we find that the phase transition is a first-order one at $\mu_{B}=0$, which is consistent with that of the pure gauge sector of QCD \cite{Boyd:1996bx,Fukushima:2010bq}. This feature is very different from that given in the previous work \cite{Li:2022erd}, where we only gave a qualitative investigation on the phase transition in a simple Einstein-dilaton-scalar system at zero chemical potential.

The paper is organized as follows. In Sec. \ref{sec-model}, we outline the holographic QCD model, and then focus on the EMDS system that will be mainly addressed. The equation of motion will be derived from the action of the model, and the boundary condition will be given for numerical calculation. In Sec. \ref{sec-phatrafinmu}, we investigate the QCD phase transition at finite $\mu_{B}$ in the EMDS system and show the numerical results. The model parameters will be fixed by fitting the two-flavor lattice results of the equation of state at $\mu_{B}=0$. We then obtain the equation of state and the chiral transition at finite $\mu_{B}$. The QCD phase diagram will also be obtained from the model. In Sec. \ref{sec-conclus}, we give a conclusion on our work with a few discussions.

\section{The improved soft-wall AdS/QCD model coupled with an EMD system}\label{sec-model}

The action of the two-flavor holographic QCD model can be divided into two parts: $S=S_G+S_M$. The part of gravitational background is an Einstein-Maxwell-dilaton (EMD) system that can be written in the string frame,
\begin{align}\label{act-grav1}
    S_G & = \frac{1}{2\kappa_5^2}\int d^5x\sqrt{-g}e^{-2\phi}\left[R -h(\phi) F_{MN}F^{MN}  
    \right.  \notag\\
   &\quad\left. +4(\partial\phi)^2 -V(\phi)\right] ,
\end{align}
where $\kappa_{5}^{2}=8\pi G_{5}$ is the effective Newton constant. The chemical potential effects can be introduced into the system by the Abelian gauge field $A_{M}$ of the action. The dilaton $\phi$ has been included to break the conformal symmetry, in order to give a sensible description for QCD phase transition \cite{Gubser:2008yx}. Later we will specify the form of the dilaton potential $V(\phi)$ and also the gauge kinetic function $h(\phi)$ which characterizes the coupling strength of the gauge field $A_{M}$. 

The flavor part of the action comes from an improved soft-wall AdS/QCD model with an additional coupling term of the Abelian gauge field and the bulk scalar field:
\begin{align}\label{act-flav1}
    S_M & = -\kappa\int d^5x\sqrt{-g}e^{-\phi}\mathrm{Tr}\Big\{|DX|^2 +V_X(X,\phi)      \nonumber \\
        & \quad +\tilde{\lambda}_{3}(\phi) F_{MN}F^{MN}|X|^{2} +\frac{1}{4g_5^2}(F_L^2+F_R^2)\Big\} ,
\end{align}
where the covariant derivative $D^MX=\p^MX -iA_L^MX+i X A_R^M$, and the chiral gauge field strengh takes the form $F_{L,R}^{MN} =\partial^MA_{L,R}^N-\partial^NA_{L,R}^M-i[A_{L,R}^M,A_{L,R}^N]$. The coupling term $\tilde{\lambda}_{3}(\phi)$ will be fixed below. The potential for the bulk scalar and the dilaton takes the form
\begin{align}\label{VX2}
    V_X(X,\phi) =m_5^2|X|^{2} -\lambda_1\phi |X|^{2} +\lambda_2|X|^{4} ,
\end{align}
where the bulk scalar mass is determined by the mass-dimension relation $m_5^2L^2 =\Delta_X(\Delta_X-4)$ with $\Delta_X=3$ being the scaling dimension of the dual operator $\bar{q}_Rq_L$ of the scalar field in the boundary \cite{Erlich:2005qh}.

The potential $V_X$ in Eq. (\ref{VX2}) has been applied to investigate thermodynamic properties of QCD in the case of $\mu_{B}=0$ \cite{Fang:2019lsz, Li:2022erd}, which shows that such a form of $V_{X}$ could provide a good description for both the equation of state and the chiral transition, at least on the qualitative level. Thus it seems natural for us to generalize to the finite chemical potential case in order to check whether this type of holographic QCD model could still produce consistent results of QCD phase transition with other models or lattice simulations. The coupling term of the Abelian gauge field and the bulk scalar field in Eq. (\ref{act-flav1}) has also been considered in previous studies with the aim to realize the correct behaviors of chiral transition at finite $\mu_{B}$ \cite{Chen:2019rez}. We remark that the role of this coupling term is in some sense like that in Ref. \cite{Chen:2019rez}, but in a rather different model setup.

The holographic QCD model is built in a gravitational background with the metric ansatz
\begin{align}\label{stringmetric}
    ds^2 &= \frac{L^2 e^{2 A_S(z)}}{z^2} \left(-f(z)dt^2 + dx^i dx^i +\frac{dz^2}{f(z)}\right) ,
\end{align}
where $L$ is the curvature radius of the asymptotic AdS$_{5}$ spacetime. Without loss of generality, we just take $L=1$ below. At finite temperature, this metric represents an asymptotic AdS black hole to be solved from the equation of motion of the system. We require $f(z_{h})=0$ at the event horizon $z_{h}$ of the black hole.

\subsection{The Einstein-Maxwell-dilaton-scalar system}

In this work, we are not going to consider the vacuum fluctuations of the matter fields, which could be neglected in comparison to the vacuum itself when tackling the problems of QCD phase transition. For simplicity, the vacuum expectation value (VEV) of the bulk scalar field is just taken to be $\langle X\rangle =\frac{\chi}{2}I_2$, with $I_2$ denoting the $2\times2$ unit matrix \cite{Erlich:2005qh}. Thus the holographic QCD model is reduced to an EMDS system:
\begin{align}\label{Eintwoscal-str1}
    S & =S_G+S_{\chi}      \nonumber                                                                          \\
      & =\frac{1}{2\kappa_5^2}\int d^5x\sqrt{-g}e^{-2\phi}\Big[R -h(\phi) F_{MN}F^{MN} \notag\\
      &\quad +4(\partial\phi)^2 -V(\phi) -\beta e^{\phi}\Big(\frac{1}{2}(\partial\chi)^2 +V(\chi,\phi)
      \notag\\
      &\quad +\frac{\tilde{\lambda}_{3}(\phi)}{2} F_{MN}F^{MN} \chi^{2}\Big)\Big] ,
\end{align}
where $\beta=16\pi G_5\kappa$ controls the coupling strength between the bulk background and the matter part, and the potential for the scalar VEV $\chi$ and the dilaton $\phi$ takes the form
\begin{align}\label{Vchi1}
    V(\chi,\phi) & =\mathrm{Tr}\,V_X(\VEV{X},\phi)        \nonumber                          \\
                 & =\frac{1}{2}(m_5^2-\lambda_1\phi)\chi^{2} +\frac{\lambda_2}{8} \chi^{4} .
\end{align}

For convenience, we usually transform to the Einstein frame by taking the metric ansatz
\begin{align}\label{einst-metric}
    ds^2 & = \frac{L^2 e^{2 A_E(z)}}{z^2} \left(-f(z)dt^2 + dx^i dx^i +\frac{dz^2}{f(z)}\right)
\end{align}
with the warp factor $A_E(z)$ related to the string-frame factor $A_S(z)$ by $A_E=A_S -\frac{2}{3}\phi$. The bulk action (\ref{Eintwoscal-str1}) in the Einstein frame can then be written as
\begin{align}\label{Eintwoscal-ef1}
    S &=\frac{1}{2\kappa_5^2}\int d^5x\sqrt{-g_{E}}\Big[R_{E} -\omega(\phi) F_{MN}F^{MN}   \nonumber \\
      &\quad -\frac{4}{3}(\partial\phi)^2 -V_E(\phi) -\beta e^{\phi}\Big(\frac{1}{2}(\partial\chi)^2 +V_E(\chi,\phi)   \nonumber \\
      &\quad +\frac{\hat{\lambda}_{3}(\phi)}{2} F_{MN}F^{MN} \chi^{2} \Big)\Big],
\end{align}
where
\begin{align}\label{Vphi-Vchi}
    \begin{split}
   \omega(\phi) &= h(\phi) e^{\frac{4\phi}{3}} ,  \\
   V_E(\phi) &= e^{\frac{4\phi}{3}}V(\phi),  \\
   V_E(\chi,\phi) &= e^{\frac{4\phi}{3}}V(\chi,\phi), \\
  \hat{\lambda}_{3}(\phi) &=\tilde{\lambda}_{3}(\phi) e^{-\frac{4}{3} \phi} .
  \end{split}
\end{align}

We could do a rescaling $\phi_c= \sqrt{8/3}\,\phi$ to convert the kinetic term of the dilaton $\phi$ to a canonical one. Following Ref. \cite{Gubser:2008yx}, we adopt a simpler form of the dilaton potential
\begin{align}\label{phi-potent1}
    V_c(\phi_c) =\frac{1}{L^2}\left(-12\cosh\gamma\phi_c +b_2\phi_c^2 +b_4\phi_c^4\right) ,
\end{align}
and take $V_E(\phi) =V_c(\phi_c)$. This guarantees the bulk geometry to have an asymptotic AdS structure near the boundary:
\begin{align}\label{phi-potent-uv1}
    V_c(\phi_c\to 0) \simeq \frac{-12}{L^2} +\frac{b_2-6\gamma^2}{L^2}\phi_c^2 +\mathcal{O}(\phi_c^4) .
\end{align}
The mass-dimension relation gives
\begin{align}\label{gamma-b2}
    b_2 =6\gamma^2 +\frac{\Delta(\Delta-4)}{2} ,
\end{align}
where $\Delta$ denotes the scaling dimension of the dual operator of the dilaton field. As shown in Ref. \cite{Li:2022erd}, the specific value of $\Delta$ in the Breitenlohner-Freedman bound actually does not affect the qualitative behaviors of phase transition. Thus we just take $\Delta=3$ for simplicity.

We adopt a form of the gauge kinetic function
\begin{align}
    \omega(\phi) =\frac{c_{0}}{4}e^{-c_{1}\phi_{c}} +\frac{1-c_{0}}{4\operatorname{sech}(c_{3}c_{4})} \operatorname{sech} \left[c_{3}\left(\phi_{c} -c_{4}\right)\right] ,
\end{align}
which approaches $1/4$ as $\phi_{c}\to 0$ in the UV limit. Such a form of $\omega(\phi)$ is inspired by those considered in Refs. \cite{DeWolfe:2010he, DeWolfe:2011ts}, which provide a suitable description for the quark susceptibility at $\mu_{B}=0$. For the last unfixed function $\hat{\lambda}_{3}(\phi)$, we will adopt a simple exponential form $\hat{\lambda}_{3}(\phi)=\lambda_{3} e^{k\phi}$ in order to produce proper chiral transition behaviors at finite $\mu_{B}$. We remark that the coupling term related to $\hat{\lambda}_{3}$ has few influences on the equation of state at finite $\mu_{B}$.

\subsection{Equation of motion and boundary condition}

By the variational method, the Einstein field equation and the equations of motion for the Abelian gauge field $A_{M}$, the dilaton $\phi$, and the scalar VEV $\chi$ can be derived from the action (\ref{Eintwoscal-ef1}) as
\begin{align}
    & R_{MN} -\frac{1}{2}g_{MN}R +\omega(\phi)\left(\frac{1}{2} g_{MN} F_{AB}F^{AB}-2F_{MA} F_{N}^{\,\,\,A}\right)      \nonumber              \\
    & +\frac{4}{3}\left(\frac{1}{2}g_{MN}\partial_J\phi\,\partial^J\phi -\partial_M\phi\partial_N\phi\right) +\frac{1}{2}g_{MN}V_E(\phi)       \nonumber \\
    & +\frac{\beta}{2} e^{\phi}\left(\frac{1}{2}g_{MN}\partial_J\chi\,\partial^J\chi -\partial_M\chi\partial_N\chi\right) +\frac{\beta}{2}g_{MN}e^{\phi}V_E(\chi,\phi)   \notag\\
    & +\frac{\beta}{2} \hat{\lambda}_{3}(\phi) e^{\phi}\left(\frac{1}{2} F_{AB} F^{AB} g_{MN}-2F_{MA} F_{N}^{\,\,\,A}\right) \chi^{2} =0 ,   \label{eins-eq0}\\
  & \nabla_{M}\left[w(\phi)F^{MN} +\frac{\beta}{2} \hat{\lambda}_3(\phi)e^{\phi}\chi^{2} F^{MN}\right] =0 ,  \label{EMDS-eomAt}\\
  & \frac{8}{3}\nabla_{M}\nabla^{M}\phi -\partial_{\phi}w(\phi)F_{MN}F^{MN} -\partial_{\phi} V_E(\phi)   \nonumber\\ & -\frac{\beta}{2}e^{\phi} g^{MN} \partial_M\chi \partial_N\chi -\beta\partial_{\phi}\left[e^{\phi}V_{E}(\chi,\phi)\right]   \nonumber\\ & -\frac{\beta}{2} \partial_{\phi}\left[\hat{\lambda}_3(\phi)e^{\phi}\right] F_{MN}F^{MN}\chi^{2} =0 ,   \label{EMDS-phieom}\\
  & \nabla_{M}\left(e^{\phi}\nabla^{M}\chi\right) -\hat{\lambda}_3(\phi) e^{\phi} F_{MN}F^{MN}\chi   \notag\\
 & -e^{\phi}\partial_{\chi}V_{E}(\chi,\phi) =0 .  \label{EMDS-chieom}
\end{align}

At finite chemical potential, we only keep the time-component $A_{t}$ of the Abelian gauge field to be nonzero. In terms of the metric ansatz (\ref{einst-metric}) and the assumption that the bulk fields depend only on the fifth-dimension coordinate $z$, Eqs. (\ref{eins-eq0}) - (\ref{EMDS-chieom}) can then be simplified to the following five indenpendent equations:
\begin{align}
   & f'' +3A_E'f' -\frac{3}{z}f' -4 z^{2} \omega(\phi) e^{-2 A_{E}} A_{t}^{\prime 2}  \notag\\
   & -2\beta z^{2} \hat{\lambda}_{3}(\phi) e^{-2 A_{E}+\phi} \chi^{2} A_{t}^{\prime 2} =0 ,   \label{fz-eom2}\\
   & A_E'' +\frac{2}{z}A_E' -A_E'^2 +\frac{4}{9}\phi'^2 +\frac{\beta}{6}e^{\phi}\chi'^2 =0 , & \label{AE-eom2}\\
  & A_{t}^{\prime\prime} +\frac{A_{t}^{\prime}}{2z e^{\phi} \omega(\phi) +\beta z \hat{\lambda}_{3}(\phi) e^{2 \phi} \chi^{2}} \Big[2\omega(\phi)e^{\phi} (zA_{E}^{\prime} -1)   \notag\\
 & +2ze^{\phi} \partial_{\phi} \omega(\phi)\phi^{\prime} +\beta z e^{2\phi} \partial_{\phi} \hat{\lambda}_{3}(\phi) \phi^{\prime} \chi^{2}   \notag\\
 & +\beta \hat{\lambda}_{3}(\phi) e^{2 \phi}\left(\chi^{2}\left(z A_{E}^{\prime} +z \phi^{\prime}-1\right) +2z\chi\chi^{\prime}\right)\Big] =0 ,  \label{At-eom2}\\
 & \phi'' +\left(3 A_E' +\frac{f'}{f}-\frac{3}{z}\right)\phi' -\frac{3\beta}{16}e^{\phi}\chi'^2 -\frac{3e^{2A_E} \partial_{\phi}V_E(\phi)}{8z^2f}  \notag\\
 & +\frac{3z^{2} e^{-2 A_{E}} A_{t}^{\prime 2} \partial_{\phi} \omega(\phi)}{4f} -\frac{3\beta e^{2A_E}\partial_{\phi}\left(e^{\phi}V_E(\chi,\phi)\right)}{8z^{2} f}  \notag\\
 & +\frac{3\beta z^{2} e^{\phi-2 A_{E}} \chi^{2} A_{t}^{\prime 2}}{8f} \left(\hat{\lambda}_{3}(\phi)+\partial_{\phi} \hat{\lambda}_{3}(\phi)\right) =0 ,  \label{dilaton-eom2}\\
  & \chi'' +\left(3A_E' +\phi' +\frac{f'}{f}-\frac{3}{z}\right)\chi' -\frac{e^{2A_E} \partial_{\chi} V_E(\chi,\phi)}{z^2 f}   \notag\\
 & +\frac{2z^{2} \hat{\lambda}_{3}(\phi) e^{-2A_{E}}}{f} A_{t}^{\prime 2} \chi =0 ,  \label{scalarvev-eom2}
\end{align}
where the function $f(z)$ and the electrostatic potential $A_{t}(z)$ are required to satisfy the following boundary conditions:
\begin{align}
   f(0) &=1, \qquad   f(z_h)=0 ,  \label{bc-fzAt1}\\
   A_{t}(0) &=\mu_{B} , \qquad 
   A_{t}(z_{h})=0 .   \label{bc-fzAt2}
\end{align}
The UV asymptotic solution for Eqs. (\ref{fz-eom2}) - (\ref{scalarvev-eom2}) can be obtained as
\begin{align}
    f(z)    & = 1 -f_4 z^4 +\cdots ,     \label{f-uv3}                                                                                                  \\
    A_E(z)  & = -\frac{1}{108}\left(3\beta m_q^2\zeta^2 +8 p_1^2\right)z^2      \nonumber                                                               \\
            & \quad -\frac{1}{24}\beta p_1 m_q^2\zeta^2 (2\lambda_1+11) z^3 +\cdots ,    \label{AE-uv3}                                                 \\
            A_{t}(z) &=\mu_{B} -\kappa_{5}^{2}n_{B}z^{2} -\frac{4\sqrt{6}}{9} p_{1} \kappa_{5}^{2}n_{B}\Big[c_{0} c_{1}   \notag\\ 
&\quad +(c_{0}-1) c_{3} \tanh (c_{3} c_{4})\Big]z^{3}+\cdots , \\
    \phi(z) & = p_1 z +\frac{3}{16} \beta m_q^2\zeta ^2 (\lambda_1+6)z^2 +p_3 z^3       \nonumber                                                       \\
            & \quad -\left[\frac{1}{48} \beta p_1 m_q^2\zeta ^2 \left(9\lambda_1^2 +111 \lambda_1 +286\right) \right.      \nonumber                    \\
            & \quad \left. -\frac{4}{9} p_1^3 \left(12 b_4-6 \gamma ^4+1\right)\right] z^3\ln z +\cdots ,  \label{phi-uv3}                               \end{align}
\begin{align}
    \chi(z) & =m_q\zeta z +p_1 m_q\zeta(\lambda_1 +5) z^2 +\frac{\sigma}{\zeta} z^3       \nonumber                                                     \\
            & \quad  -\left[\frac{1}{96} m_q^3\zeta^3 \left(\beta  \left(9\lambda_1^2+108 \lambda_1 +308\right)-24 \lambda_2\right) \right.   \nonumber \\  &\quad \left. +\frac{1}{18}p_1^2m_q\zeta \left(9\lambda_1^2 +111 \lambda_1+286\right) \right]z^3\ln z +\cdots ,   \label{chi-uv3}
\end{align}
where $\zeta=\frac{\sqrt{3}}{2 \pi}$ is a normalization constant \cite{Cherman:2008eh}, $m_q$ denotes the current quark mass, $\sigma$ the chiral condensate, and $n_{B}$ the baryon number density. From the UV expansions (\ref{f-uv3}) - (\ref{chi-uv3}), we read another two conditions:
\begin{align}\label{bc-phichi}
    \phi^{\prime}(0)=p_{1} , \qquad \chi^{\prime}(0) =m_{q}\zeta .
\end{align}
In terms of the boundary conditions (\ref{bc-fzAt1}), (\ref{bc-fzAt2}) and (\ref{bc-phichi}), the bulk fields can be solved numerically from Eqs. (\ref{fz-eom2}) - (\ref{scalarvev-eom2}). The baryon number density $n_{B}$ and the chiral condensate $\sigma$ can then be extracted from the UV asymptotics of $A_{t}$ and $\chi$, respectively.

\section{QCD phase transition at finite chemical potential}\label{sec-phatrafinmu}

\subsection{Equation of state and chiral transition}

Once we could solve the EMDS system, the equations of state and the chiral transition can be investigated simultaneously at finite $\mu_{B}$. The temperature and the entropy density are given by the formulae:
\begin{align}\label{temperentro}
    T=\frac{|f'(z_h)|}{4\pi} , \qquad  s =\frac{2\pi e^{3A_E(z_h)}}{\kappa_5^2 z_h^3} .
\end{align}
The pressure $p$ can be calculated by the differential relation
\begin{align} \label{pTdiffrelat}
    dp=s dT +n_{q}d\mu_{q} ,
\end{align}
and the energy density can then be obtained from the thermodynamic relation $\varepsilon =-p+sT+\mu_{q}n_{q}$, with the quark number density $n_{q} =3n_{B}$ and the quark chemical potential $\mu_{q}=\mu_{B}/3$.

To do the numerical calculation, we need to fix the model parameters, which could be determined by fitting the lattice results of the equation of state in the two-flavor case \cite{Allton:2003vx,Karsch:2001vs}. First, we set the background-matter coupling $\beta=1$ and the quark mass $m_{q}=5\MeV$. By fitting the equation of state at $\mu_{q}=0$, we can fix the parameters $G_{5}=0.582$, $\gamma=0.75$, $b_{4}=0.02$ and $p_{1}=0.473$. Other parameters of the model are related to the chemical potential effects and the properties of chiral transition. For instance, the values of the coefficients $c_{0}$, $c_{1}$, $c_{3}$, $c_{4}$ and $\lambda_{3}$, $k$ have direct effects on the equation of state at finite $\mu_{q}$. In order to fix these parameters, we could use the quark susceptibility defined by
\begin{align}\label{chiq-def}
    \chi_{2}^{q} =\frac{\partial^{2} \left(P / T^{4}\right)}{\partial\left(\mu_{q} / T\right)^{2}} =\frac{\partial\left(n_{q} / T^{3}\right)}{\partial\left(\mu_{q} / T\right)} .
\end{align}
The values of $\lambda_{1}$, $\lambda_{2}$, $\lambda_{3}$, and $k$ are crucial for a proper realization of the chiral transition behaviors at finite $\mu_{q}$. Since the background fields and the matter fields are coupled with each other, the thermodynamic properties at finite $\mu_{q}$ actually depend on all the parameters of the model. By a global fitting, we can fix the remaining parameters to be $c_{0}=0.725$, $c_{1}=30$, $c_{3}=0.47$, $c_{4}=0.45$, $\lambda_{1}=-1$, $\lambda_{2}=10$, $\lambda_{3}=-0.65$, and $k=-1.67$.

\begin{figure}
    \begin{center}
\includegraphics[width=63mm,clip=true,keepaspectratio=true]{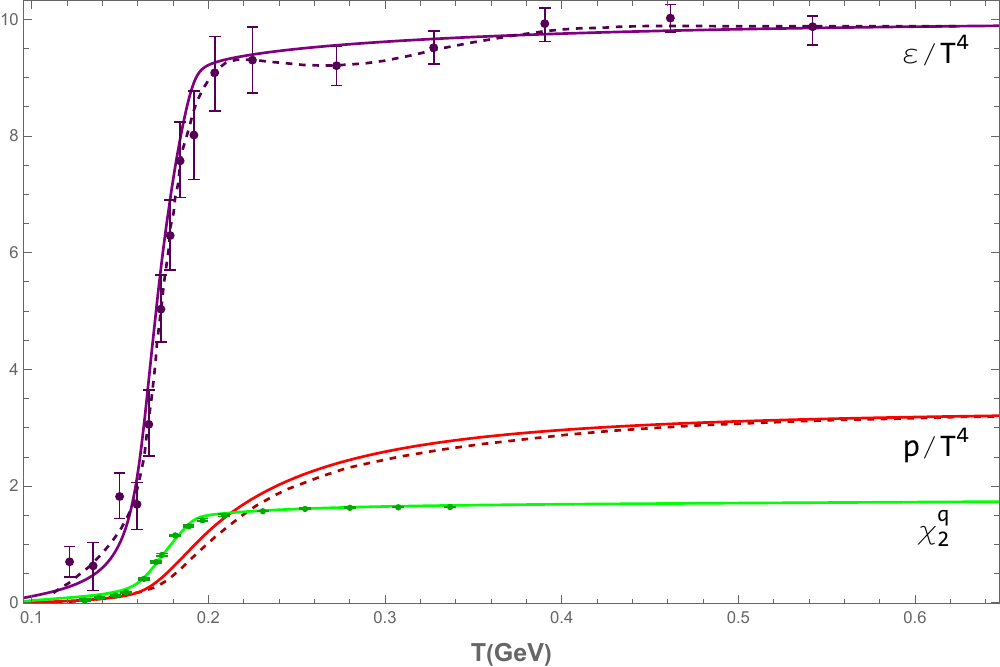}
    \end{center}
    \caption{Comparison of the model results for the scaled energy density $\epsilon/T^{4}$, the pressure $p/T^{4}$ and the quark susceptibility $\chi_{2}^{q}$ with the lattice results at $\mu_{q}=0$. The model results are denoted by solid lines, while the lattice data are denoted by dashed lines or points with error bars \cite{Allton:2003vx,Karsch:2001vs}.}
    \label{fig-epchi2qmu0}
\end{figure}

\begin{figure}
    \begin{center}
\includegraphics[width=68mm,clip=true,keepaspectratio=true]{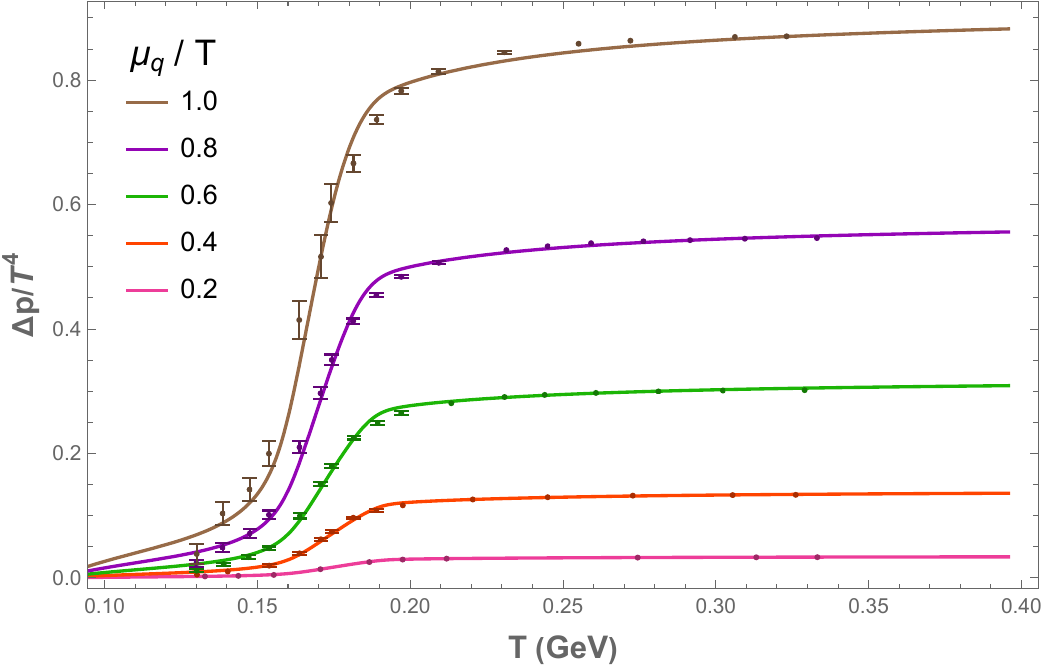}
        \vskip 0.5cm\hspace*{-0.1cm}
\includegraphics[width=68mm,clip=true,keepaspectratio=true]{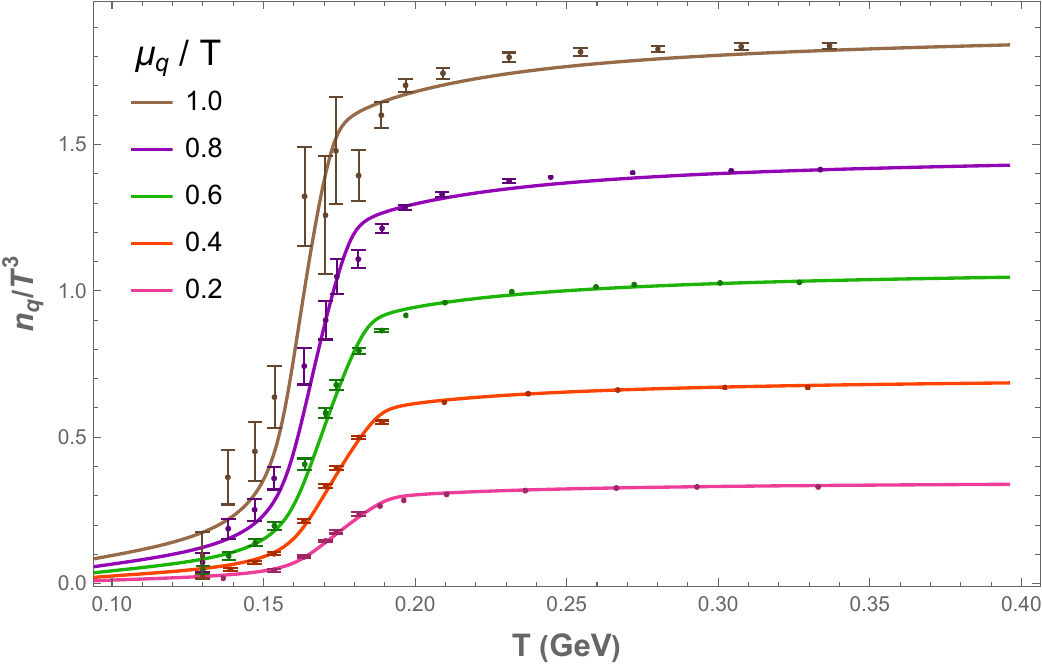} \vskip 0.3cm 
    \end{center}
    \caption{Comparison of the model results for the scaled pressure $\Delta p/T^{4}$ and the quark number density $n_{q}/T^{3}$ with the lattice results for $\mu_{q}/T=0.2, 0.4, 0.6, 0.8, 1$. The lattice data are taken from Ref. \cite{Allton:2003vx,Allton:2005gk}.}
    \label{fig-pnqTmubTc}
\end{figure}

In Fig. \ref{fig-epchi2qmu0}, we present the model calculations for the scaled energy density $\epsilon/T^{4}$, the pressure $p/T^{4}$ and the quark susceptibility $\chi_{q}/T^{2}$ at $\mu_{q}=0$, which have been used to fix the parameters of the model. We can see that the model results have a good match with the lattice results in the two-flavor case \cite{Allton:2003vx,Karsch:2001vs}. Then we can apply our model to investigate the properties of the equation of state and the chiral transition at nonzero chemical potential. In Fig. \ref{fig-pnqTmubTc}, we present the model results for the scaled pressure $\Delta p/T^{4}$ and the quark number density $n_{q}/T^{3}$ with fixed values of $\mu_{q}/T$, which are compared with the lattice results in the two-flavor case. We find that the model results are in good agreement with the lattice data.

\begin{figure}
    \begin{center}
\includegraphics[width=65mm,clip=true,keepaspectratio=true]{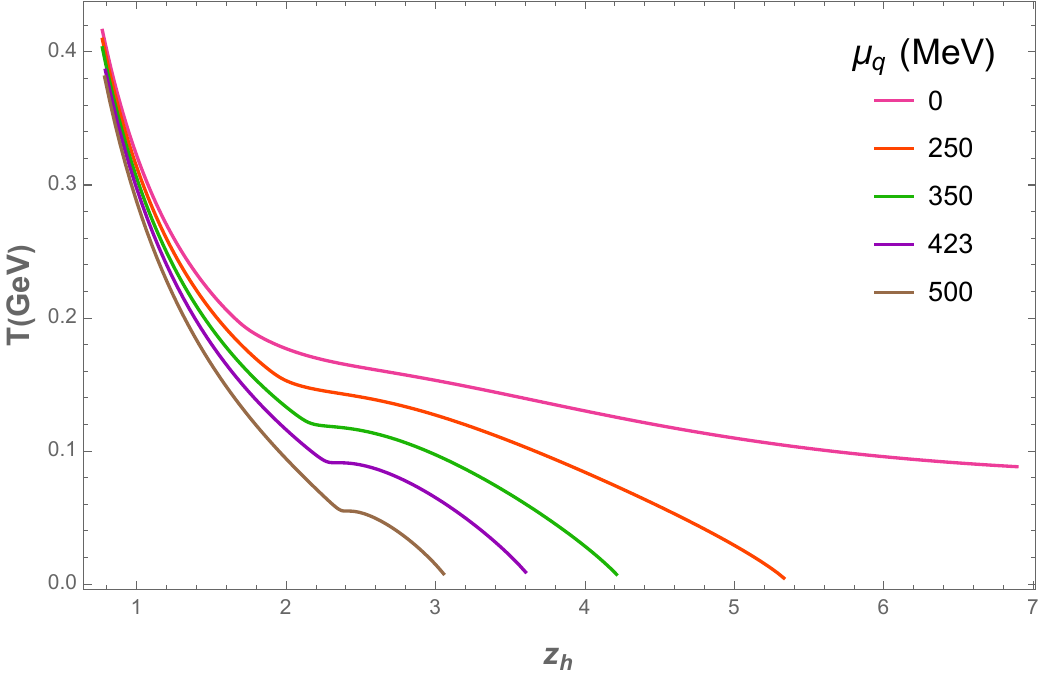}
    \end{center}
    \caption{The temperature $T$ as a function of the horizon $z_{h}$ at $\mu_{q} =0, 250, 350, 423, 500 \MeV$.}
    \label{fig-Tzhmuqc}
\end{figure}

\begin{figure}
    \begin{center}
\includegraphics[width=65mm,clip=true,keepaspectratio=true]{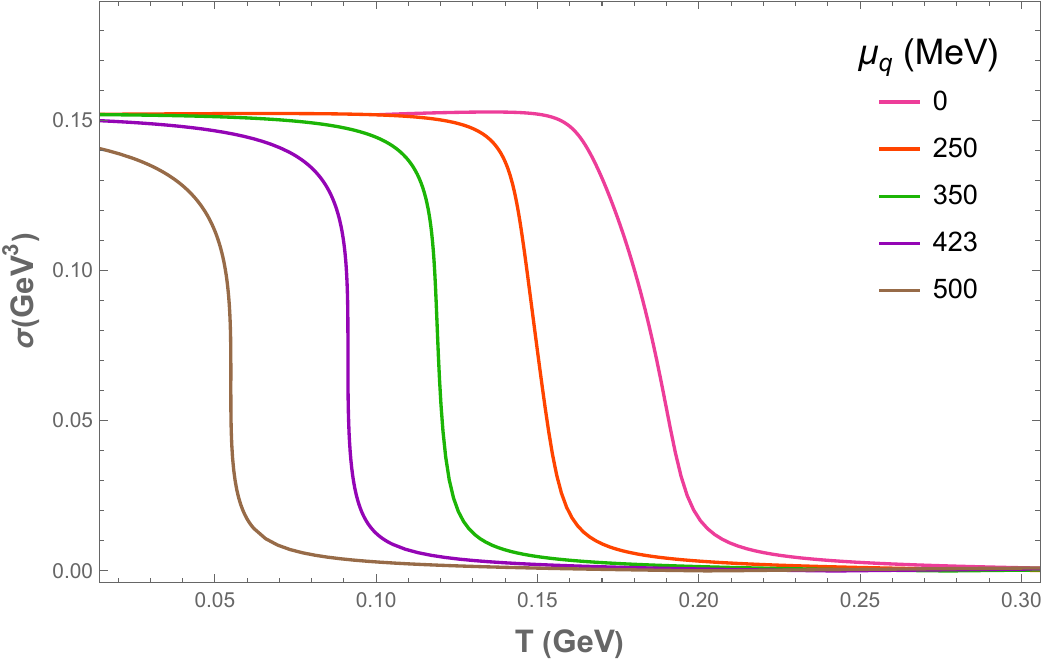}
    \end{center}
    \caption{The behaviors of the chiral condensate $\sigma$ with respect to the temperature $T$ at $\mu_{q} =0, 250, 350, 423, 500 \MeV$.}
    \label{fig-sigzhmuqc}
\end{figure}

We also investigate the behaviors of the equation of state and the chiral transition with fixed values of $\mu_{q}$. We present the temperature $T$ as a function of the horizon $z_{h}$ at five different values of $\mu_{q}$ in Fig. \ref{fig-Tzhmuqc}, from which we find that for smaller values of $\mu_{q}$, the temperature decreases monotonically with the event horizon. However, when $\mu_{q}$ increases beyond a critical point $\mu_{q}^{E}\simeq 423\MeV$, the behavior of $T$ with respect to $z_{h}$ will become nonmonotonic in a small region of $z_{h}$, which signals that the transition order of the thermodynamic quantities obtained from the model will be changed around the critical value $\mu_{q}^{E}$. The chiral transition behaviors with fixed $\mu_{q}$ are presented in Fig. \ref{fig-sigzhmuqc}, from which we can see that the chiral transition is a smooth crossover at smaller values of $\mu_{q}$, yet when $\mu_{q}$ becomes larger and larger, the chiral condensate $\sigma$ descends more and more rapidly with the increase of temperature $T$ in the transition region, until $\mu_{q}$ reaches the critical point $\mu_{q}^{E}$, at which the chiral transition becomes a second-order phase transition, and then it converts to a first-order phase transition when $\mu_{q}$ is larger than $\mu_{q}^{E}$. These behaviors of chiral transition with respect to the chemical potential are consistent with that of the $T-z_{h}$ curves in Fig. \ref{fig-Tzhmuqc}.

\begin{figure}
    \begin{center}
\includegraphics[width=66mm,clip=true,keepaspectratio=true]{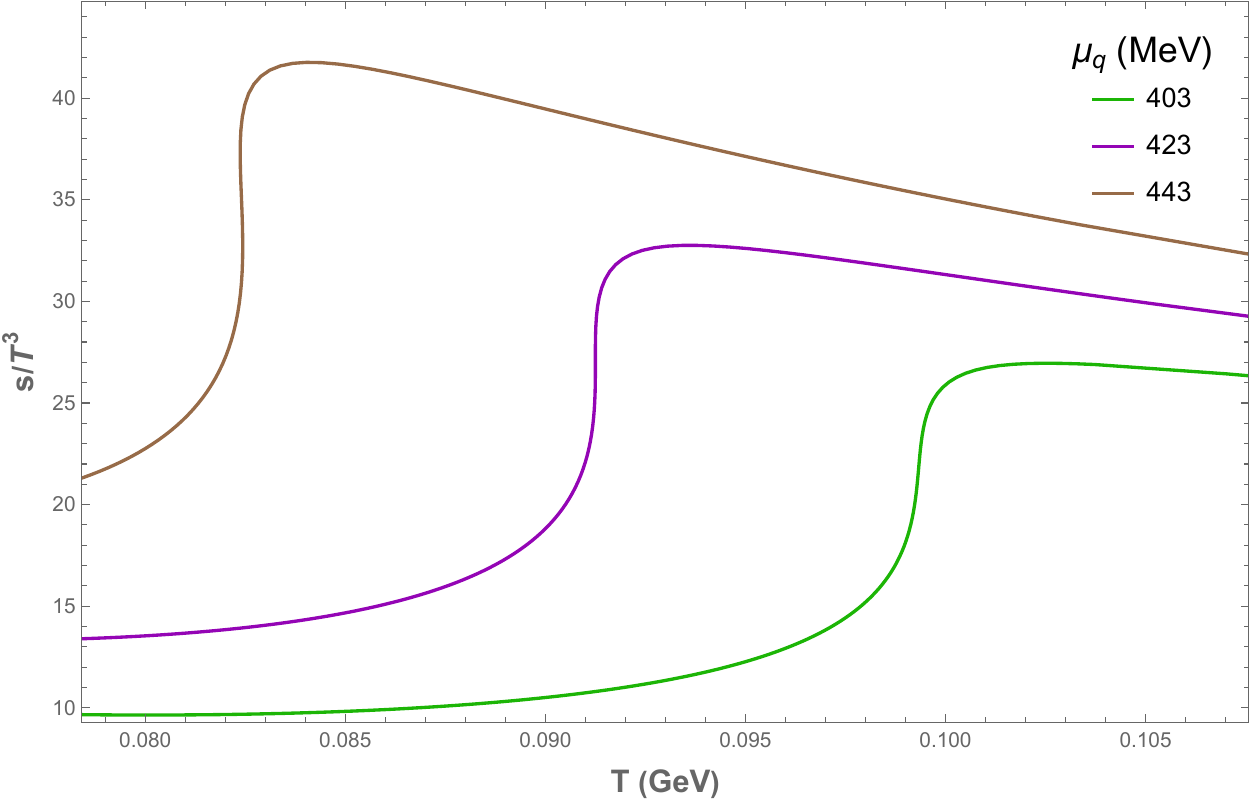}
        \vskip 0.5cm\hspace*{-0.1cm}
\includegraphics[width=68mm,clip=true,keepaspectratio=true]{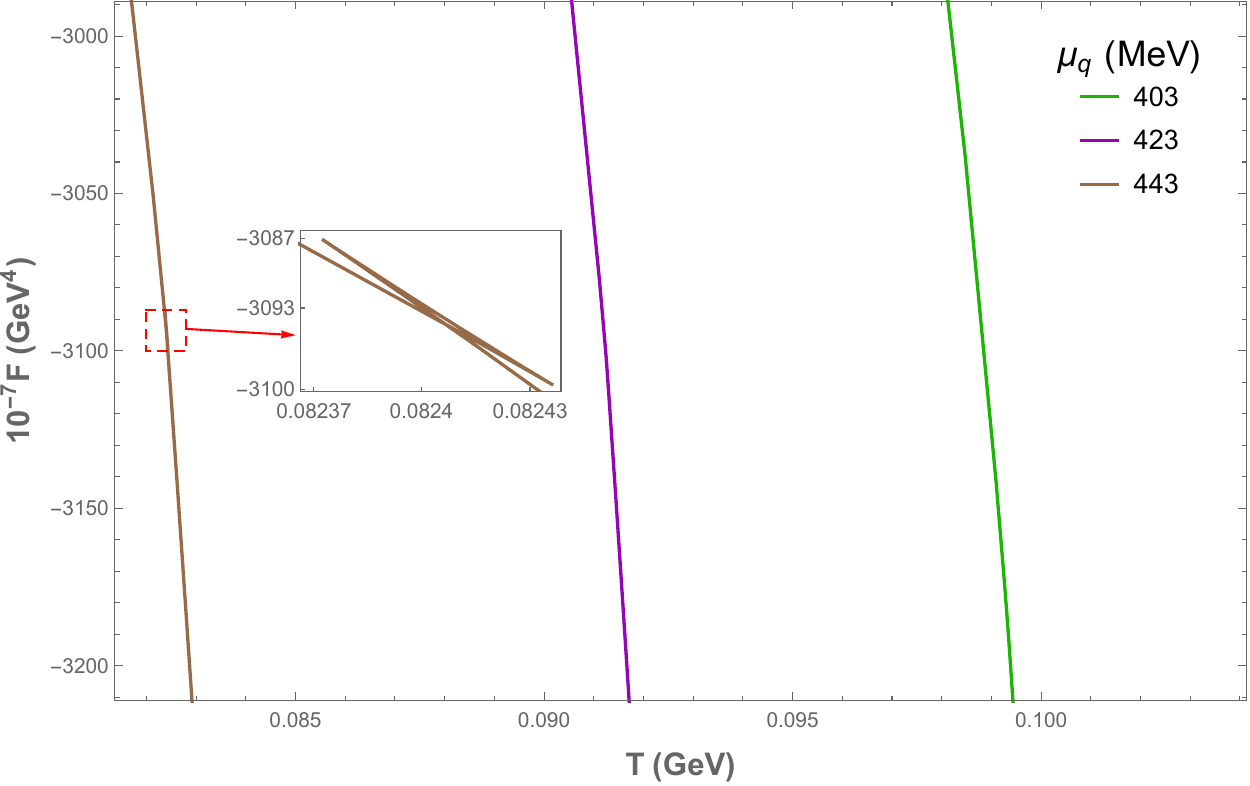} \vskip 0.3cm 
    \end{center}
    \caption{The behaviors of the scaled entropy density $s/T^{3}$ and the free energy $F$ around the transition temperature for $\mu_{q}=403, 423, 443 \MeV$.}
    \label{fig-sFT3difmuq}
\end{figure}

Note that the change of the transition order is not that obvious from the chiral transition behaviors shown in Fig. \ref{fig-sigzhmuqc}. We then calculate the scaled entropy density $s/T^{3}$ and the free energy $F=-p$ around the critical point $\mu_{q}^{E}$. In Fig. \ref{fig-sFT3difmuq}, we present the model results for three neighboring values of $\mu_{q}$, from which we can see clearly that the free-energy line displays a swallow-tail shape at $\mu_{q}=443\MeV$, which is a clear signal for a first-order phase transition. The intersection point of the free-energy line determines the critical temperature of the first-order transition.

\subsection{QCD phase diagram}

We can obtain the QCD phase diagram as long as we could determine the transition temperature at each chemical potential, which can be attained from the analysis of the equation of state and the chiral transition. As mentioned above, the critical temperature $T_{c}$ for the first-order phase transition at large $\mu_{q}$ can be determined by the intersection point of the free-energy line. As $\mu_{q}$ decreases to $\mu_{q}^{E}\simeq 423\MeV$, the swallow-tail shape of the free-energy line disappears, with the critical temperature $T^{E}\simeq 91.2\MeV$, at which a second-order phase transition occurs. The phase transition turns into a smooth crossover below $\mu_{q}^{E}$. There is no unique way to determine the transition temperature $T_{c}$ for the crossover case. We may use the minimum of the speed of sound $c_{s}$, which is a suitable probe to characterize the drastic change of degrees of freedom between the confinement phase of hadrons and the deconfinement phase of quark-gluon plasma. In addition, we may also use the maximally decreasing point of the chiral condensate $\sigma$ to signify the chiral transition temperature.

\begin{figure}
    \begin{center}
\includegraphics[width=68mm,clip=true,keepaspectratio=true]{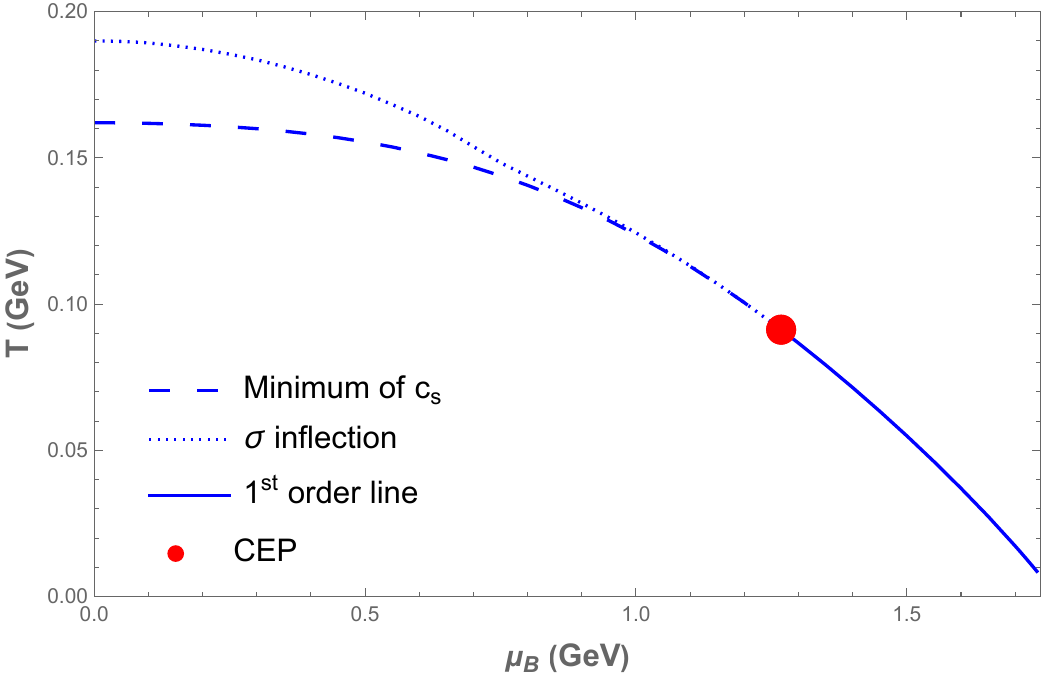}
    \end{center}
    \caption{The QCD phase diagram obtained from the model. The minimum of the speed of sound $c_{s}$ and the maximally decreasing point of the chiral condensate $\sigma$ are denoted by the dashed and dotted lines, respectively. The first-order transition line is denoted by solid line. The CEP is marked by a red point.}
    \label{fig-phasdiag}
\end{figure}

The model prediction for the QCD phase diagram is presented in Fig. \ref{fig-phasdiag}, where we use the baryon chemical potential $\mu_{B}$, instead of the quark chemical potential $\mu_{q}$. Note that there is a CEP located at $(\mu_{B}^{E}, T^{E}) =(1268\MeV, 91.2\MeV)$, which links the first-order transition line in the range of $\mu_{B} >\mu_{B}^{E}$ with the crossover transition line in the range of $\mu_{B}<\mu_{B}^{E}$. It seems that the critical point $\mu_{B}^{E}$ of our model with two flavors is larger than those obtained from other models including a recent holographic study on the phase structure of two-flavor QCD \cite{Sasaki:2010jz,Fu:2019hdw,Gao:2020qsj,Gao:2016qkh,Cai:2022omk}. It can be seen from Fig. \ref{fig-phasdiag} that the two crossover lines given by the minimum of $c_{s}$ and the $\sigma$ inflection do not coincide, with the chiral transition temperature greater than the deconfinement transition temperature at smaller values of $\mu_{B}$, and they converge gradually to the CEP with the increase of $\mu_{B}$. At $\mu_{B}=0$, the transition temperature determined by the minimum of $c_{s}$ is about $162\MeV$, while the chiral transition temperature is about $190\MeV$.

\subsection{Pure gauge sector}

With suitable choice of parameters, the EMDS system with the action (\ref{Eintwoscal-str1}) is capable of describing both the equation of state and the chiral transition at finite chemical potential, and the properties of phase transition obtained from the model are consistent with the lattice results and the common expectations. Now we would like to decouple the matter part with the bulk background by setting $\beta=0$. It is interesting to see how the equation of state and other thermodynamic quantities behave in the reduced EMD system which may holographically characterize the gluon dynamics of QCD.

\begin{figure}
    \begin{center}
\includegraphics[width=68mm,clip=true,keepaspectratio=true]{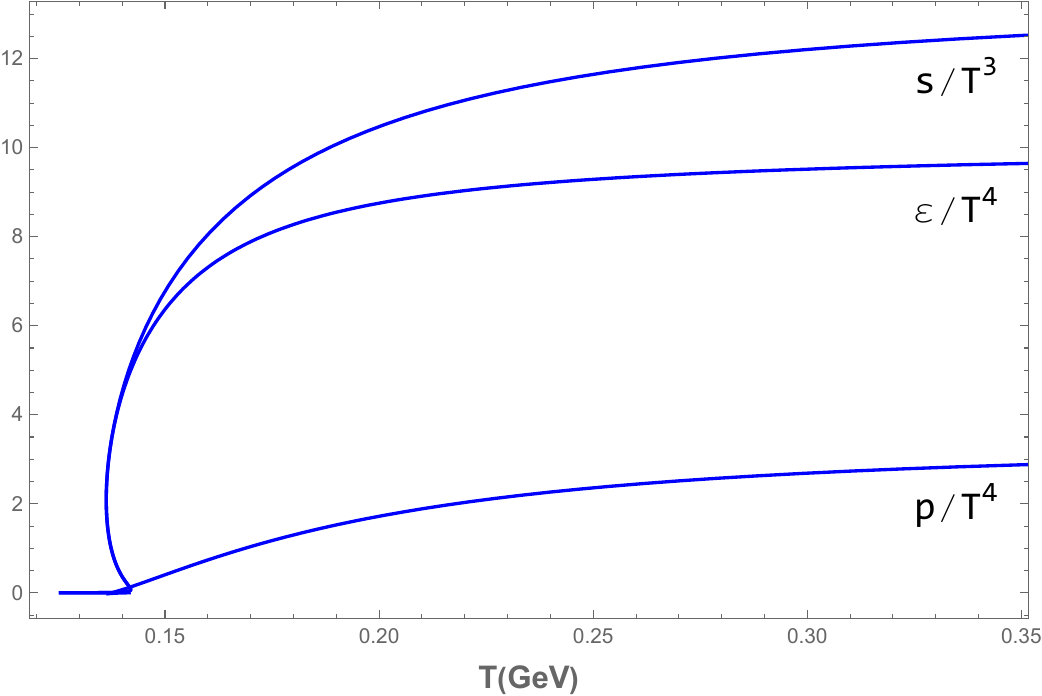}
    \end{center}
    \caption{The scaled entropy density $s/T^{3}$, the energy density $\epsilon/T^{4}$ and the pressure $p/T^{4}$ at $\mu_{q}=0$ in the case of $\beta=0$.}
    \label{fig-sepspFbet01}
\end{figure}

\begin{figure}
    \begin{center}
\includegraphics[width=68mm,clip=true,keepaspectratio=true]{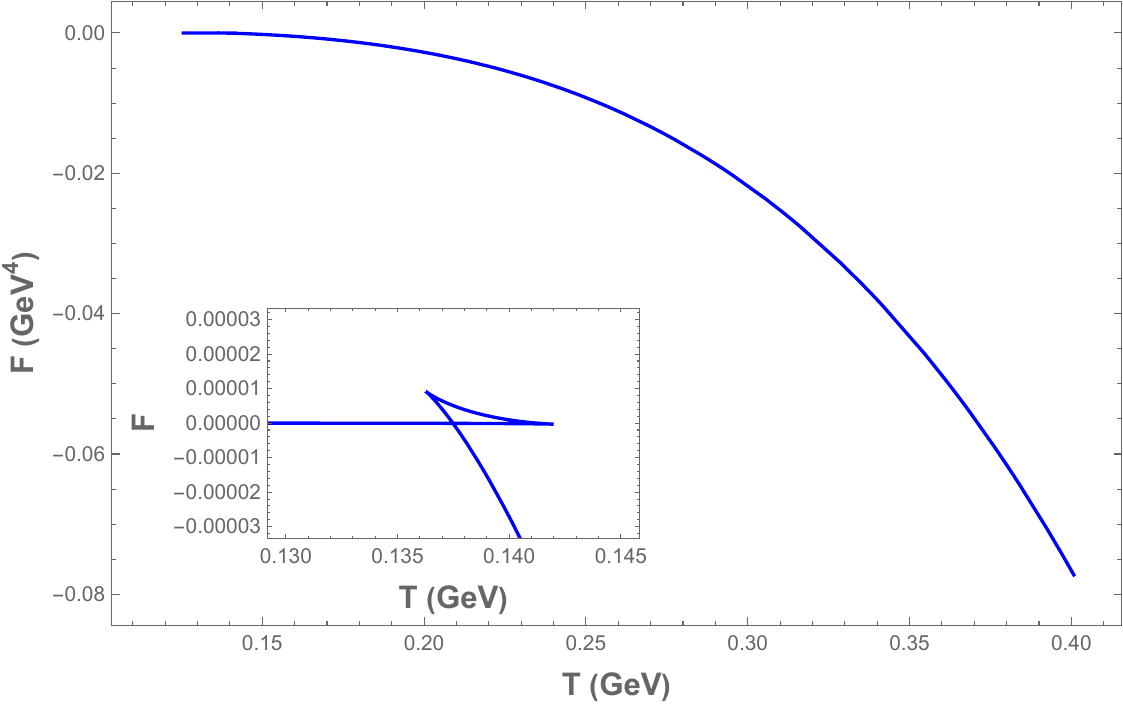}
    \end{center}
    \caption{The free energy $F$ at $\mu_{q}=0$ in the case of $\beta=0$.}
    \label{fig-sepspFbet02}
\end{figure}

We calculate the scaled entropy density $s/T^{3}$, the energy density $\epsilon/T^{4}$ and the pressure $p/T^{4}$ at zero chemical potential in this EMD system with the fixed parameters. The numerical results are presented in Fig. \ref{fig-sepspFbet01}. Clearly, the EMD system of this model exhibits a first-order phase transition at $\mu_{q}=0$, which is qualitatively consistent with the lattice simulations for pure Yang-Mills theory \cite{Boyd:1996bx,Fukushima:2010bq}. The free-energy line in this case has been shown in Fig. \ref{fig-sepspFbet02}, from which we can read the critical temperature $T_{c}\simeq 137.5 \MeV$.

\section{Conclusion and discussion}\label{sec-conclus}

We investigated the properties of phase transition and phase diagram in an improved soft-wall AdS/QCD model coupled with an EMD system in the two-flavor case. With fixed parameters, this model yields the equation of state in quantitative agreement with the lattice results of two flavors at both zero and nonzero chemical potentials. It also generates sensible chiral transition behaviors consistent with the equation of state at finite chemical potential. Note that the deconfinement and chiral transitions are interrelated with each other, and occur simultaneously in the first-order transition region of large $\mu_{B}$. This is a natural result of the fact that the bulk background fields and the vacuum scalar field have been coupled together. But in the crossover region, because of different order parameters being used, the transition temperatures obtained from them are also different. In addition, we find that the first-order phase transition beyond $\mu_{B}^{E}$ is actually very weak in our model, as can be seen by the free-energy lines in Fig. \ref{fig-sFT3difmuq}. 

The QCD phase diagram in the $T-\mu_{B}$ plane has been obtained, with a CEP located at $(\mu_{B}^{E}, T^{E}) =(1268\MeV, 91.2\MeV)$. The large value of the critical point $\mu_{B}^{E}$ in our model forms an obvious contrast to those obtained from the previous studies \cite{Sasaki:2010jz,Fu:2019hdw,Gao:2020qsj,Gao:2016qkh}. By the equation of state, we also find that the reduced EMD system of our model generates a first-order phase transition at zero chemical potential, which is consistent with that of the pure gauge theory \cite{Boyd:1996bx,Fukushima:2010bq}. To sum up, all these features indicate that the phase structure of our model is quite similar to that of QCD, at least on a qualitative level, as long as $\mu_{B}$ is not too large. We may wonder whether a more quatitative description for the phase structure of QCD could be obtained from such a holographic framework.

For a convincing description of the QCD phase transition, we actually need to give the information of hadron spectrum. It remains to be seen whether the mass spectra of light hadrons could be reproduced in this holographic QCD model. It will be interesting to generalize the two-flavor investigations of this work to the $2+1$ flavor case by introducing another vacuum scalar field related to the strange flavor, in which case we may have other ways to obtain the QCD phase diagram containing a CEP. With the rise of the astronomy of neutron stars and gravitational waves, it will also be interesting to apply this holographic QCD model to investigate the equation of state and other properties of neutron stars.

\section*{Acknowledgements}
This work is supported by the National Natural Science Foundation of China (NSFC) under Grant No. 11905055, the Natural Science Foundation of Hunan Province, China under Grants No. 2020JJ5026 and No. 2023JJ30115, and the Fundamental Research Funds for the Central Universities.

\bibliography{refs-AdSQCD}

\end{document}